# Fractals of graphene quantum dots in photoluminescence of shungite

B.S. Razbirin[1], N.N Rozhkova[2], E.F. Sheka[3*], D.K. Nelson[1], A.N. Starukhin[1], A.S.Goryunov[4]

[1]*Ioffe Physical-Technical Institute, RAS, Saint Petersburg, Russia*
[2]*Institute of Geology, Karelian Research Centre RAS, Petrozavodsk*
[3]*Peoples' Friendship University of Russia, Moscow*
[4]*Institute of Biology Karelian Research Centre RAS, Petrozavodsk, Russia*

E-mail: sheka@icp.ac.ru

## Abstract

Photoluminescence of graphene quantum dots (GQDs) of shungite, attributed to individual fragments of reduced graphene oxide (rGO), has been studied for the frozen rGO colloidal dispersions in water, carbon tetrachloride, and toluene. Morphological study shows a steady trend of GQDs to form fractals and a drastic change in the colloids fractal structure caused by solvent was reliably established. Spectral study reveals a dual character of emitting centers: individual GQDs are responsible for the spectra position while fractal structure of GQD colloids provides high broadening of the spectra due to structural inhomogeneity of the colloidal dispersions and a peculiar dependence on excitation wavelength. For the first time, photoluminescence spectra of individual GQDs were observed in frozen toluene dispersions which pave the way for a theoretical treatment of GQD photonics.

## 1 Introduction

Originally, the term 'graphene quantum dot' (GQD) appeared in theoretical researches and was attributed to fragments limited in size, or domains, of a single-layer two-dimensional graphene crystal. The subject of the investigations concerned the quantum size effects, manifested in the spin [1, 2], electronic [3] and optical [4-9] properties of the fragments. The latter study significantly stimulated the interest in GQD and their attractive applications (see, for example, [10] and references therein) so that the question arose of their preparation. This proved to be a difficult task and the progress achieved by now has been presented in exhaustive reviews [11, 12]. On the basis of spectral studies, have found that, in almost all cases, the GQDs are not single-layer graphene domains, but multi-layer formations containing up to 10 layers of reduced graphene oxide (rGO) of less than 30 nm in size.

Optical spectroscopy, photoluminescence (PL), in particular, was the primary method of studying the properties of the GQDs. The review [12] presents a complete picture of the results, which can be presented by the following set.

1. Regardless of the method of obtaining GQDs, the final product is a mixture of particles differing in size - both width and thickness.
2. Morphology of GQDs revealed that the particle size is not determined by the production process, although depending on the starting materials. The average linear dimension is about 10 nm, while the maximum is 60 nm. The average thickness of particles indicates multi-layer, five-layer, in most cases, GQDs while obtaining single-layer GQDs is also not uncommon (see, for example [13-17]).
3. Fourier transform infrared spectroscopy and photoelectron spectra show that in almost all the cases studied, the chemical composition of GQDs corresponds to partially oxidized graphene.

4. The absorption spectra in the visible and UV range show a well-marked size effect that is manifested as a red shift of the spectrum with increasing the GQDs size.

Detailed description of these features with the presentation of their possible explanations and links to the relevant publications is given in the review [12].

As seen from the synopsis, optical spectroscopy of GQDs gives a complicated picture with many features. However, in spite of this diversity, common patterns can be identified that can be the basis of the GQDs spectral analysis, regardless of method of their preparation. These general characteristics characterizing GQDs include: 1) structural inhomogeneity of GQDs solutions, better called dispersions; 2) low concentration limit that provides surveillance of the PL spectra; 3) dependence of the GQD PL spectrum on the solvent, and 4) dependence of the GQD PL spectrum on the excitation light wavelength. It is these four circumstances that determine usual conditions under which the spectral analysis of complex polyatomic molecules is performed. Optimization of conditions, including primarily the choice of solvent and the experiments performance at low temperature, in many cases, led to good results, based on structural PL spectra (see, for example, the relevant research of fullerenes solutions [18-21]). In this paper, we will show that implementation of this optimization for spectral analysis of the GQDs turns out quite successful.

## 2 Graphene Quantum Dots of Natural Origin

Synthetic GQDs described in the previous section have recently been complemented with GQDs of natural origin [22, 23]. As shown, GQDs present the main structural peculiarity of shungite of the Karelian deposits. Basing on the detailed analysis of physical and chemical properties of shungite and its derivation, was established [22], that this mineral, which must be regarded as one of the natural carbon allotropes, is multistage fractal nets of rGO fragments of $\leq 1$ nm in size. The generality of the basic structural elements of shungite and synthetic GQDs as well as the ability of the former to disperse in water provided a basis of a research project [23] that was aimed at establishing the generality of the spectral properties of aqueous dispersions of shungite and synthetic GQDs and proving the structural formula of shungite given above. The conducted spectral studies provided the desired confirmation exhibiting at the same time particular features of the observed spectral characteristics that allow one to penetrate into the depth of the structural and spectral peculiarities of the GQDs dissolved in different solvents.

## 3 Fractal Nature of the Object under Study

A concept on GQD evidently implies a dispersed state of a number of nanosize rGO fragments. Empirically, the state is provided by the fragments dissolution in a solvent. Once dissolved, the fragments unavoidably aggregate forming colloidal dispersions. As mentioned earlier, so far only aqueous dispersions of synthetic GQDs have been studied [11, 12]. In the case of shungite GQDs, two molecular solvent, namely, carbon tetrachloride and toluene were used when replacing water in the pristine dispersions. In each of these cases, the colloidal aggregates are the main object of the study. In spite of that so far there has not been any direct confirmation of their fractal structure there are serious reasons to suppose that it is an obvious reality. Actually, first, the fragments formation occurred under conditions that unavoidably involve elements of randomness in the course of both laboratory chemical reactions and natural graphitization [22]; the latter concerns the fragments size and shape. Second, the fragments structure certainly bears the stamp of polymers, for which fractal structure of aggregates in dilute dispersions has been convincingly proven (see [24] and references therein).

As shown in [24], the fractal structure of colloidal aggregates is highly sensitive to the solvent around, the temperature of the aggregates formation, as well as other external actions such as mechanical stress and so forth. This fact makes the definition of quantum dots of colloidal dispersions at the structural level quite undefined. In the case of the GQDs of different origin, the situation is additionally complicated since the aggregation of synthetic (Sy) and shungite (Sh) rGO fragments occurred under different external conditions. In view of this, it must be assumed that rGO-Sy and rGO-Sh aggregates of not only different, but the same solvent dispersions are quite different.

Looking for the answer to the question if the same term GQD can be attributed to colloidal dispersion in the above two cases, one should recall that the feature of fractal structures is that fractals are typically self-similar patterns, where "self-similar" means that they are "the same from near as from far" [25]. This means that the peculiarities of, say, optical behaviour of each of the two rGO-Sy and rGO-Sh colloidal dispersions obey the same law. From this viewpoint, there apparently is no difference which structural element of a multilevel fractal structure of their colloidal aggregates should be attributed to a quantum dot. However, the identity of both final and intermediate fractal structures of aggregates in different solvents is highly questionable and only the basic rGO structural units cast no doubts. Because of this, GQDs of both rGO-Sy ant rGO-Sh dispersions should to associate with rGO individual fragments. Therefore, different fractal nets of GQDs provided by different colloidal dispersions present the object under the current study. Addressing to spectral behavior of the dispersions, we should expect an obvious generality provided by the common nature of GQDs, but simultaneously complicated by the difference in fractal packing of the dots in the different-solvent dispersions. The latter study concerns mainly the rGO-Sh dispersions [22] that will be considered in detail below.

## 4 rGO-Sh Aqueous Dispersions

In full agreement with commonly used methods for the preparation of colloidal dispersions of graphene and its derivatives [26, 27], rGO-Sh aqueous dispersions were obtained by sonication of the pristine shungite powder for 15 min with an ultrasonic disperser UZ-2M (at a frequency of 22 kHz and the operating power 300 W) followed by filtration and ultracentrifugation [28]. The maximum achievable concentration of carbon is less than 0.1 mg / ml, which is consistent with poor water solubility of graphene and its derivatives [26]. The resulting dispersions are quite stable, and their properties vary little during the time. The size-distribution characteristic profile of rGO-Sh aggregates is shown in Fig. 1a. As can be seen from the figure, the average size of the aggregates is 54 nm, whereas the distribution is quite broad - 26 nm so that the resulting colloids are significantly inhomogeneous. The inhomogeneity obviously concerns both size and shape (and, consequently, chemical composition) of basic rGO fragments and, consequently, GQDs. The structure of the carbon condensate formed after water evaporation from the dispersion droplets on a glass substrate is shown in Fig. 1b and 1c. As seen from the figure, the condensate is of fractal structure formed by aggregates, the shape of which is close to spherical. It should be mentioned that the condensate fractal structure should not be identical to that one of the pristine dispersion [24], although, no doubt, some continuity of the structure should take place.

Figure 2a shows an overview on the characteristic patterns of the emission spectrum of rGO-Sh aqueous dispersions at different excitation $\lambda_{exc}$ at 80 K. Increasing the temperature to 293 K does not cause a significant change in the spectra, resulting in only a slight broadening of their structural component related to the Raman spectrum (RS) of water at the fundamental frequency of the O-H stretching vibrations of ~3400 cm$^{-1}$. When excited at $\lambda_{exc}$ 405 and 457 nm, RS superimposes the broad luminescence band in the region of 17000-22000 cm$^{-1}$. The emission of the dispersion at $\lambda_{exc}$ $\lambda_{exc}$ 532 nm is weak. Figure 2b shows the PL spectra of the aqueous dispersion at $\lambda_{exc}$ 405 and 457 nm after subtraction of the RS of water. Both spectra are broad, bell-shaped distributions characteristic of the PL spectra of rGO-Sy aqueous dispersions (see [12]). In spite of the large width of PL spectra, their position in the same spectral region for both rGO-Sy and rGO-Sh aqueous dispersions evidences a common nature of emitting GQDs.

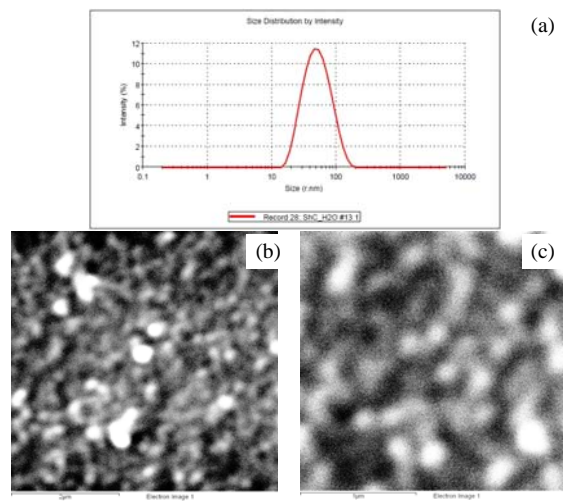

**Figure 1.** Size-distribution profile of colloidal aggregates in shungite water dispersion (a) and SEM images of the dispersion condensate on glass substrate (b) and (c): Scale bar 2 μm and 1 μm, respectively. Carbon concentrations constitute 0.1 mg/ml.

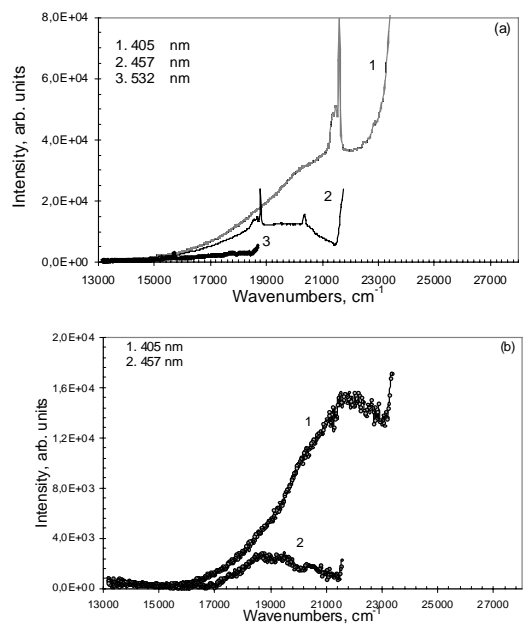

**Figure 2.** Photoluminescence spectra of shungite water dispersions at 80K as observed (a) and after subtraction of Raman scattering of water (b). Figures mark excitation wavelengths.

The similarity of the spectral behavior of the two dispersions is also spread over a considerable overlapping of their absorption and PL spectra so that a set of new PL spectra can be excited with an increase $\lambda_{exc}$ within practically each PL spectrum. Of course, a shift of these new PL spectra maxima towards longer wavelengths with increasing $\lambda_{exc}$ is observed. Such a behavior usually indicates the presence of the inhomogeneously broadened absorption spectrum of the emitter, which widely overlap with PL one and whose excitation at different $\lambda_{exc}$ within the overlapping region results in selective excitation of different sets of emitting centers. In the case of rGO-Sy dispersions, the spectra inhomogeneous broadening is usually explained by scatter in the GQDs (rGO fragments) linear dimensions [12]. However, not only GQDs linear dimensions, but their shape as well as the composition of colloidal aggregates may largely vary, which should be expected for the rGO-Sh dispersions, in particular. This is clearly seen on the example of various aggregates structure of the condensates shown in Fig. 1b and 1c. Unfortunately, large width of PL spectra does not allow one to exhibit those spectral details that might speak about aggregated structure of GQDs.

## 5. rGO-Sh Dispersions in Organic Solvents

Traditionally, the best way to overcome difficulties caused by inhomogeneous broadening of optical spectra of complex molecules is the use of their dispersions in frozen crystalline matrices. The choice of solvent is highly important. Thus, the water is a "bad" solvent since the absorption and emission spectra of dissolved large organic molecules usually are broadband and unstructured. In contrast, frozen solutions of complex organic molecules, including, say, fullerenes [18-21], in carbon tetrachloride or toluene, in some cases provide a reliable monitoring of fine-structured spectra of individual molecules (Shpolskii's effect [29]). Detection of PL structural spectra or structural components of broad PL spectra not only simplify spectral analysis but indicate the dispersing of emitting centers into individual molecules. It is this fact that was the basis of the solvent choice when studying spectral properties of shungite GQDs [23].

Organic rGO-Sh dispersions were prepared from the pristine aqueous dispersions in the course of sequential replacement of water by isopropyl alcohol first and then by carbon tetrachloride or toluene [30]. The morphology and spectral properties of these dispersions turned out to be different, thereby consider them separately.

### 5.1. rGO-Sh Dispersions in Carbon Tetrachloride

When analyzing CTC-dispersions morphology, a drastic change in the size-distribution profiles of the dispersions aggregates in comparison with that one of the aqueous dispersions (see Fig. 3) was the first highly important result. The finding directly evidences the fractal nature of the GQDs aggregates in water since only under this condition so strong effect caused by the solvent substitution can be observed [24]. The second result concerns the high incertitude in this fractal structure. Thus, Figs 3a and 3b present size-distribution profiles related to CTC-dispersions of most different by this parameter. Figures 3c and 3d show images of agglomerates of films obtained when drying the CTC-dispersions droplets on glass. As can be seen by comparing Figs. 1 and Fig. 3, the average colloidal aggregate size is increased when water is substituted by CTC. Simultaneously, increased scatter of sizes that is comparable with the size itself in the limiting case. The nearly spherical shape of aggregates in Fig. 1b and 1c is replaced by lamellar faceting, mostly characteristic of microcrystals. Noteworthy is the absence of small aggregates, which indicates a complete absence of individual GQDs in the dispersions. Therefore, the change in size-distribution profiles as well as in shape of the aggregates of the condensate evidences a strong influence of solvent on the aggregates structure thus decisively confirming their fractal character.

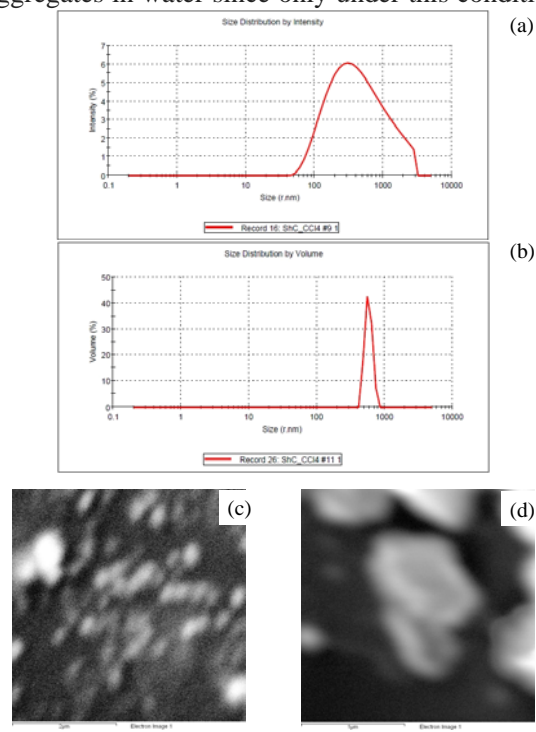

**Figure 3.** Size-distribution profiles of colloidal aggregates in shungite dispersion in carbon tetrachloride (a and b) (see text) and SEM images of the dispersion condensates on glass substrate (c) and (d): Scale bar 2 μm and 1 μm, respectively.

The conducted spectral studies are well consistent with these findings. Figure 4 shows PL spectra of CTC-dispersion DC1, morphological properties of which are near to those shown in Fig. 3a. The dispersions have a faint yellow-brown color, which indicates the presence of significant absorption of the solutions in the visible region (see Fig. 4a). PL spectra were studied for a wide range of dispersions obtained at different time. As found, both these spectra behavior at different exciting lasers $\lambda_{exc}$ and the PL spectra shape of different dispersions are largely similar while the spectra intensity can differ substantially. Arrows in Fig. 4a show wave number values $\lambda_{exc}^{-1}$ corresponding to laser lines at 405, 457, 476.5, 496.5, 514.5 and 532 nm. As seen in Fig. 4a, the dispersion absorption increases when advancing to the UV region. It can be assumed that the absorption of each component of the aggregates conglomerate increases with decreasing $\lambda_{exc}$, so that the excitation with UV light at $\lambda_{exc} = 337.1$ nm affects almost all the centers of luminescence in the crystal matrix. Actually, the UV excited PL spectrum in Fig. 4a is very broad and covers the region from 27000 to 15000 cm$^{-1}$. In this case, the PL spectrum overlaps with the absorption spectrum over the entire spectral range.

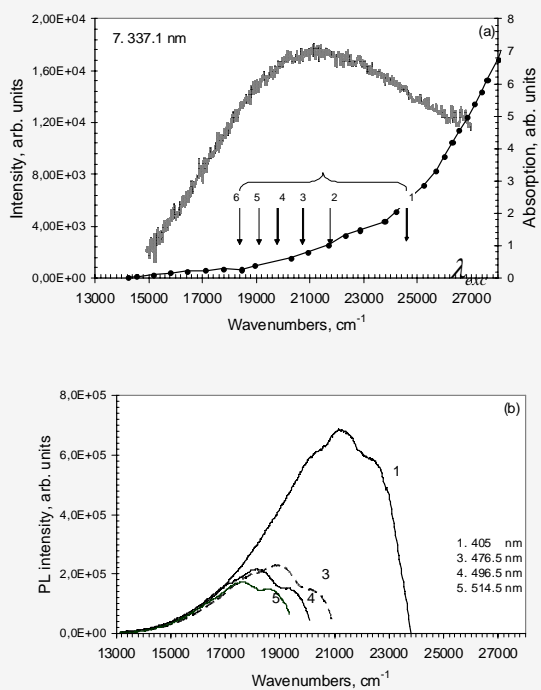

**Figure 4.** Photoluminescence (a and b) and absorption (a) spectra of shungite dispersion in carbon tetrachloride DC1 at 80K after background emission subtraction. Figures mark excitation wavelengths.

Such a large overlapping evidences the inhomogeneously broadened character of both spectra that is, the formation of an ensemble of emitting centers, which differ in the probability of emission (absorption) at given wavelength (it is assumed that the probability of energy migration between the centers is low). Indeed, successive PL excitation by laser lines 1, 3, 4 and 5 (see Fig. 4a) causes a significant modification of the PL spectra (Fig. 4b). The width of the spectra decreases as $\lambda_{exc}$ increases, the PL band maximum is shifted to longer wavelengths, and the spectrum intensity decreases. This is due to selective excitation of a certain group of centers. In general, the observed pattern is typical for structurally disordered systems discussed in the previous section. To simplify further comparative analysis of the spectra obtained at different $\lambda_{exc}$, we shall denote them according to the excitation wavelength, namely: 405-, 476-, 496-spectrum, etc.

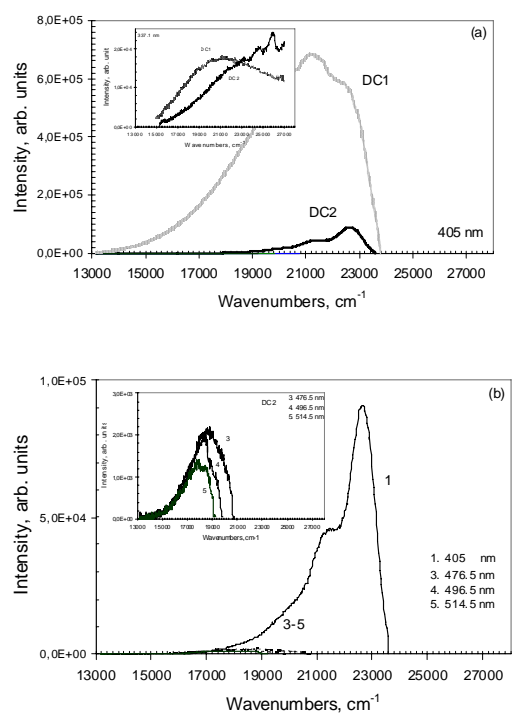

**Figure 5.** Photoluminescence spectra of shungite dispersions in carbon tetrachloride at 80K after background emission subtraction. A comparative view of 405- and 337-spectra (insert) of DC1 and DC2 dispersions (a); the same for spectra of DC2 dispersions at different excitations (b). Figures mark excitation wavelengths.

Comparing the PL spectra of dispersion DC1 at different excitations, note that 1) PL spectra obtained when excited in the region of overlapping of absorption and emission spectra in Fig. 4a, have more distinct structure than the 337-one but still evidencing a superpositioning character of the spectra; 2) intensity of the 405-spectrum is almost an order of magnitude higher than the intensity of the rest of the spectra. Before discussing the observed spectra features, let us have a look at the PL spectra of dispersion DC2 that is close in morphology to the dispersion shown in Fig. 3b. Figure 5a compares the 337- and 405- spectra of DC2 with those of DC1 described earlier. The 337-spectrum of DC2 exhibits a new UV band, intensity of its 405-spectrum decreases by several times. The spectrum of DC2 still retains a three-peak shape, but their intensities are significantly redistributed. This evidently tells about a superpositioning character of the spectrum as said before. It is important to note that the 405-spectrum of DC2 is still the most intense among other DCQ spectra (see Fig. 5b).

Comparative analysis of the PL spectra of dispersions DC1 and DC2 shows that the above-mentioned spectral regularities are sensitive to the CTC-dispersions structure and are directly related to the degree of structural inhomogeneity. Thus, the narrowing of the size-distribution profile related to dispersion DC2, undoubtedly causes narrowing of inhomogeneously broadened absorption and emission spectra, so that the intensity of the long-wavelength emission spectra of DC2 dispersion decreases. Because of the cutoff of the long-wavelength absorption spectrum of DC2 dispersion the structure of its 405-spectrum becomes more noticeable due to, apparently, additional feature of the distribution of emitting centers in DC2 over energy. Unchanged in both sets of spectra is the predominance intensity of 405-spectrum.

The difference in the structural inhomogeneity of dispersions raises the question of their temporal stability. Spectral analysis of their PL allows answering this question. In Fig. 6, we are back to dispersion DC1, but after 1.5 years (dispersion DC1*). As can be seen from Fig. 6a, in the PL spectrum DC1* there are new emitting centers, responsible for the PL in the UV region. Otherwise, 337-spectrum changes little, keeping its intensity and large width. Changes in PL spectra in the visible range are less pronounced (see Fig. 6b). Attention is drawn to high intensity of the 457-spectrum of DC1*.

Thus, the set of PL spectra obtained for rGO-Sh CTC-dispersions allows one to make the following conclusions.

1. In the low-temperature PL spectra of crystalline CTC-dispersions none of fine-structured spectra similar to Shpolskii's spectra of organic molecules was observed. This is consistent with the absence of small particles in the size-distribution profiles of the relevant colloidal aggregates.

**Figure 6.** Photoluminescence spectra of shungite dispersions in carbon tetrachloride at 80K after

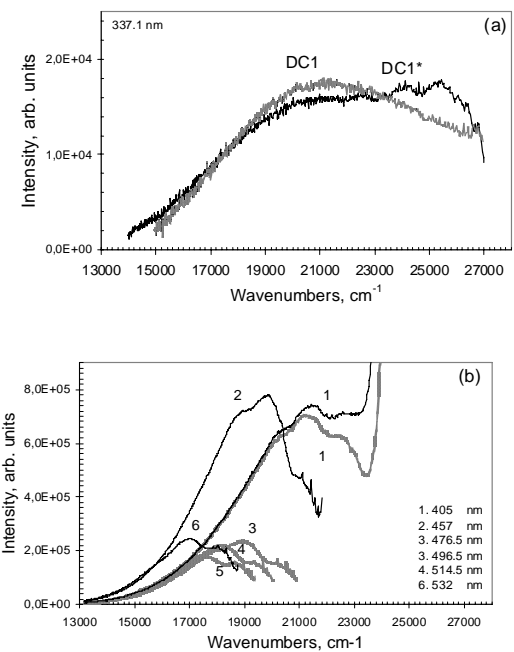

background emission subtraction. A comparative view of 337-spectra (insert) of DC1 and DC1* dispersions (see text) (a); the same for spectra of DC1 (thickened curves) and DC* (thin curves) dispersions at different excitations (b). Figures mark excitation wavelengths.

2. The PL spectra are broad and overlapping with the absorption spectrum over a wide spectral range. This fact testifies to the inhomogeneous broadening of the spectra, which is the result of non-uniform distribution of the dispersions colloidal aggregates, confirmed by morphological measurements.
3. The observed high sensitivity of PL spectra to the structural inhomogeneity of dispersions allows the use fluorescent spectral analysis as a method of tracking the process of the formation of primary dispersions and their aging over time.
4. Selective excitation of emission spectra by different laser lines allows decomposing the total spectrum into components corresponding to the excitation of different groups of emitting centers. In this case, common to all the studied dispersions is the high intensity of the emission spectra excited at $\lambda_{exc}$ 405 and 457 nm.

### 5.2. rGO-Sh Dispersion in Toluene

The behavior of toluene rGO-Sh dispersions is more intricate from both morphological and spectral viewpoints. Basic GQDs of aqueous dispersions are awfully little soluble in toluene, thereby resulting toluene dispersions are essentially colorless due to low concentration of the solute. In addition, the low concentration makes the dispersion very sensitive to any change in both the content and structure of dispersions. This causes structural instability of dispersions which is manifested, in particular, in the time dependence of the relevant size-distribution profiles. Thus, the three-peak distribution of the initial toluene dispersion shown in Fig. 7a, is gradually replaced by a single-peaked distribution in Fig. 7b for one to two hours. The last distribution does not change with time and represents the distribution of the solute in the supernatant.

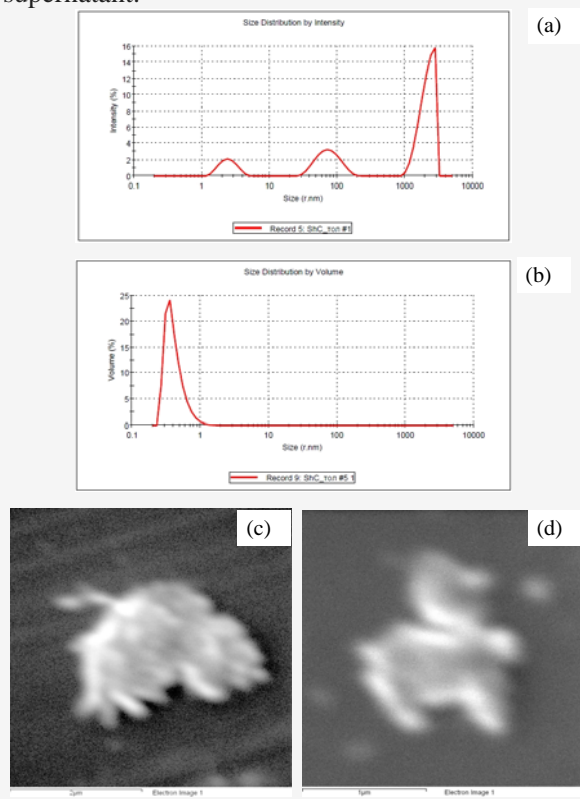

**Figure 7.** Size-distribution profiles of colloidal aggregates in shungite toluene dispersion (a and b) (see text) and SEM images of the dispersion condensates on glass substrate (c) and (d): Scale bar 2 μm and 1 μm, respectively. Carbon concentrations constitute 0.08 (a) and 0.04 (b) mg/ml .

By analogy with carbon tetrachloride, toluene causes a drastic change in the colloidal aggregates structure thus proving once again the fractal structure of the pristine GQD colloids in the aqueous dispersions. However, if the carbon tetrachloride action can be attributed to the consolidation of the pristine colloids, the toluene results in quite opposite effect leading to their dispersing. Three-peak structure in Fig. 7a shows that, at the initial stage of water replacement by toluene, in resulting liquid medium there are three kinds of particles with average linear dimensions of about 2.5, 70 and 1100 nm. All the three sets are characterized by a wide dispersion. Large particles are seen in the electron microscope (Fig. 7c and 7d) as freaky sprawled fragments. Over time, these three entities are replaced by one with an average size of ~ 0.25 nm. Thus, freshly produced dispersions containing GQD aggregates of varying complexity, turns into the dispersion of individual GQDs. It should be noted that the obtained average size seems to be too small. This might be due to that the program processing of the particle distribution in Zetasizer Nano ZS (Malvern Instruments) is based on spherical 3D particles approximation, whereby the output data can be assigned to the two-dimensional structural anisotropic particles with a big stretch. This makes to accept the value of 0.25 nm as very approximate and only consistent on the order of magnitude with the empirical value of ~1nm for the average size of GQDs in shungite [22].

This conversion of the aqueous dispersion of aggregated GQDs into the colloidal dispersion of individual GQDs in toluene is a peculiar manifestation of the interaction of solvents with fractals described in [24]. Apparently, GQD fractals are differently 'opaque' or 'transparent' with respect to CTC and toluene, which causes so big effect. Certainly, the finding may stimulate the consideration of nanosize graphene dispersions in the framework of the fractal science similarly to the polymer study [24]. As for the graphene photonics, the obtained toluene dispersion has provided investigation of individual GQDs for the first time.

Figure 8 shows the PL spectra of colloidal dispersions of individual GQDs in toluene. The *brutto* experimental spectra, each of which is a superposition of the Raman spectrum of toluene and PL spectrum of the dispersion, are presented in Fig.8a. Note the clearly visible enhancement of Raman scattering of toluene in the 20000-17000 cm$^{-1}$ region. Figure 8b shows the PL spectra after subtracting Raman spectra. The spectra presented in the figure can be divided into three groups. The first group includes the 337-spectrum (7) that in the UV region is the PL spectrum, similar in shape to the UV PL spectrum of toluene, but shifted to longer wavelengths. This part of the spectrum should apparently be attributed to the PL of some impurities in toluene. The main contribution into the PL 337-spectrum in the region of 24000-17000 cm$^{-1}$ is associated with the emission of all GQDs available in the dispersion. This spectrum is broad and structureless, which apparently indicates the structural inhomogeneity of the GQD colloids.

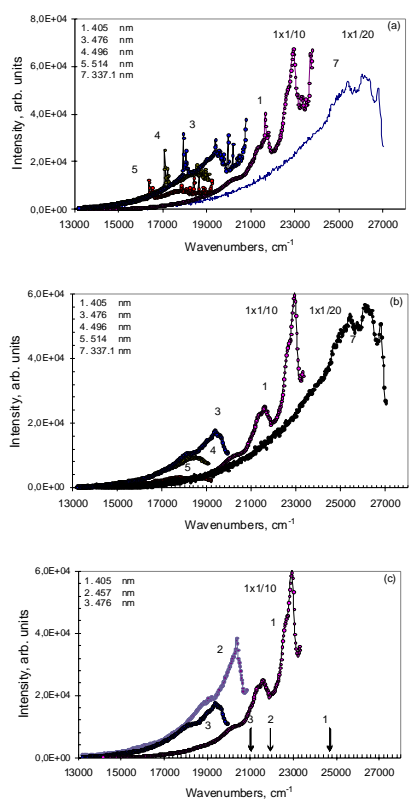

**Figure 8.** Photoluminescence spectra of shungite toluene dispersion at 80K as observed (a), after subtraction of Raman scattering of water (b), and attributed to individual GQDs only (c). Figures and arrows mark excitation wavelengths.

PL 405- and 476-spectra (1 and 3) in the region of 23000-17000 cm$^{-1}$ should be attributed to the second group. Both spectra have clearly defined structure that is most clearly expressed in the 405-spectrum. The spectrum is characteristic of a complex molecule with allowed electronic transitions. Assuming that the maximum frequency at 22910 cm$^{-1}$ determines the position of pure electronic transition, the longer wavelength doublet at ~ 21560-21330 cm$^{-1}$ can be interpreted as vibronic transitions. The distance between the doublet peaks and the pure electronic band constitutes 1350-1580 cm$^{-1}$ that is consistent with the frequencies of totally symmetric vibrations of C-C graphene skeleton, commonly observed in the Raman spectra. Similarly, two peaks of the much less intensive 476-spectrum, which are wider than in the previous case, are divided by the average frequency of 1490 cm$^{-1}$. PL 457-spectrum, shown in Fig. 8c (curve 2) is similar to spectra 1 and 3, in intensity closer to the 405-spectrum. All the three spectra are related to individuals rGO fragments albeit of different size that increases when going from 405-spectrum to 457- and 476-spectrum.

The shape of 496- and 514-spectrum substantially differs from that of the second group spectra. Instead of the two peaks observed there a broad band is observed in both cases. This feature makes these spectra attribute to the third group and associate them with the appearance of not individual frozen GQDs but with their possible clusters (such as, say, dimeric homo- (GQD+GQD) and hetero- (GQD+toluene) structured charge transfer complexes and so forth). The evidence of such a possibility will be discussed in the next section.

The conducted spectral studies of the rGO-Sh toluene dispersions confirmed once again the status of toluene as a good solvent and a good crystalline matrix, which allows for obtaining structured spectra of individual complex molecules under conditions when in other solvents the molecules form fractals. This ability of toluene allowed for the first time to get the spectra of both individual GQDs and their small clusters. The finding represents the first reliable empirical basis for further theoretical treatment of the spectra observed. Simultaneously with this, use of toluene and carbon tetrachloride as solvents convincingly showed a strong tendency of nanoscale graphenes to form fractals, which should be taken into account in practical applications.

## 6. Discussion

As follows from the results presented in the previous sections, rGO-Sh dispersions are colloidal dispersions regardless of the solvent, whether water, carbon tetrachloride or toluene. The dispersion colloids structure depends on the solvent and thereafter is substantially different. This issue deserves a special investigation. Thus, the replacement of water with carbon tetrachloride leads to multiple enlargement of the pristine colloids which promotes the formation of a quasi-crystalline image of the condensate structure. At present, the colloid detailed structure remains unclear. In contrast to carbon tetrachloride, toluene causes the decomposition of pristine colloids into individual rGO fragments. The last facts cast doubt on the possible direct link between the structure of the dispersions fractals and the elements of fractal structure of solid shungite or its post-treated condensate. The observed solvent-stimulated structural transformation is a consequence of the geometric peculiarities of fractals behavior in liquids [24]. The resulting spectral data can be the basis for further study of this effect.

The spectral behavior of the aqueous and CTC-dispersions with large colloids is quite similar, despite the significant difference in size and structure of the latter. Moreover, the features of the PL spectra of these dispersions practically replicate patterns that are typical for the aqueous rGO-Sy dispersions discussed in detail in Section 1. This allows one to conclude that one and the same structural element of the colloidal aggregates of both rGO-Sh dispersions and rGO-Sy one is responsible for the emission in spite of pronounced morphological difference of its packing in all these cases. According to the modern view on the shungite structure [22] and a common opinion on the origin of synthetic GQDs [11, 12], rGO fragments of ~1nm should play the role thus representing GQDs of the rGO colloidal dispersions in all the cases.

Specific effects of toluene, which caused the decomposition of pristine particles into individual rGO fragments with succeeding embedding them into a crystalline matrix of toluene, allowed for the first time to obtain the PL spectrum of individual rGO fragments. Obviously, resulting fragments are of different size and shape, which determines the structural inhomogeneity of toluene dispersions. This feature of toluene dispersions is common with the other dispersions and explains the dependence of PL spectra on $\lambda_{exc}$ that is the main spectral feature of GQDs, both synthetic [11, 12] and of shungite origin.

The structural inhomogeneity of GQDs colloidal dispersions is mainly caused by two reasons, namely, internal and external. Internal reason concerns the uncertainty in the structure (size and shape) of the basic rGO fragments. It is the most significant for shungite while, under laboratory conditions, the rGO fragments structure might be more standardized [11, 12]. Nanosize rGO basic structural elements of solid shungite are formed under the conditions of a serious competition of different processes [22], among which the most valuable are: 1) natural graphitization of carbon sediments, accompanied by a simultaneous oxidation of the graphene fragments and their reduction in water vapor; 2) the retention of water molecules in space between fragments and going out the water molecules from the space into the environment, and 3) the multilevel aggregation of rGO fragments providing the formation of a monolithic fractal structure. Naturally, that achieved balance between the kinetically-different-factor processes is significantly influenced by random effects, so that the rGO fragments of natural shungite, which survived during a Natural selection, are statistically averaged over a wide range of fragments that differ in size, shape, and chemical composition.

Obviously, the reverse procedure of the shungite dispersing in water is statistically also nonuniform with respect to colloidal aggregates so that there is a strong dependence of the dispersions on the technological protocol, which results in a change in the dispersion composition caused by slight protocol violations. This, in a sense, a kinetic instability of dispersing, is the reason that the composition of colloidal aggregates can vary when water is displaced by other solvent. The conducted spectral studies have confirmed these assumptions.

External reason is due to fractal structure of colloidal aggregates. The fractals themselves are highly inhomogeneous, moreover, they strongly depend on the solvent. The two reasons determine the feature of the GQD spectra in aqueous and CTC-dispersions while the first one dominates in the case of toluene dispersions. In view of this, photonics of GQDs has two faces, one of which is of graphene nature while the other concerns fractal packing of graphenes. As follows from the presented in the paper, spectra study is quite efficient in exhibiting this duality.

Thus, the structural PL spectra allow putting the question of identifying the interaction effect of dissolved rGO fragments with each other and with the solvent. Nanosize rGO fragments have high donor-and-acceptor properties (low ionization potential and high electron affinity) and can exhibit both donor and acceptor properties so that clusters of fragments (dimers, trimers, and so forth) are typical charge transfer complexes. Besides this, toluene is a good electron donor, due to which it can form a charge-transfer complex with any rGO fragment, acting as an electron acceptor. The spectrum of electron-hole states of the complex, which depends on the distance between the molecules, on the initial parameters is similar to the electron-hole spectrum of clusters of fullerenes $C_{60}$ themselves and with toluene [19], positioned by the energy in the region of 20000-17000 $cm^{-1}$. By analogy with nanophotonics of fullerene $C_{60}$ solutions [19], the enhancement of the RS of toluene is due to superposition of the spectrum over the spectrum of electron-hole states, which follows from the theory of light amplification caused by nonlinear optical phenomena [32]. Additionally, the formation of rGO-toluene charge transfer complexes may promote the formation of stable chemical composites in the course of photochemical reactions [33] that might be responsible for the PL third-group spectra observed in toluene dispersions. Certainly, this assumption requires further theoretical and experimental investigation.

## 7. Conclusion

Photonics of shungite colloidal dispersions faces the problem that big statistical inhomogeneity inherent in the quantum dot as an object of the study makes it difficult to interpret the results in details. Consequently, most important become common patterns that are observed on the background of this inhomogeneity. In case of the considered dispersions, the common patterns include, primarily, the dispersion PL in the visible region, which is characteristic for large molecules consisting of fused benzenoid rings. This made it possible to confirm the earlier findings that graphene-like structures of limited size, namely, rGO fragments are the basic structural elements for all the dispersions. The second feature concerns the dependence of the position and intensity of selective PL spectra on the exciting light wavelength $\lambda_{exc}$. This feature lies in the fact that regardless of the composition and solvent of dispersions the PL excitation at $\lambda_{exc}$ 405 and 457 nm provides the highest PL intensity while excitation at either longer or shorter wavelengths produces a much lesser intensity of the emission. The answer to this question must be sought in the calculated absorption and photoluminescence spectra of graphene quantum dots, which we attributed to nanoscale fragments of reduced graphene.

*Remarks to Experimental Techniques*

Morphological investigation of rGO-Sh dispersions concerned the definition of the size distribution of the dispersions colloidal aggregates and obtaining overview pictures of structure of films obtained therefrom by evaporation of the solvent. The relevant size-distribution profiles were obtained by dynamic light scattering by using nanoparticle size analyzer Zetasizer Nano ZS (Malvern Instruments). Processing of the results was carried out in the approximation of a spherical shape of the aggregates [34]. Spectral studies were carried out at 293 K and 80 K. Emission spectra of liquid and frozen dispersions were excited by laser lines $\lambda_{exc}$ of 337, 405, 457, 476, 496, 514 and 532 nm. The resulting spectra were normalized to the power of the laser radiation at different excitations. The spectra were recorded under identical conditions on a DFS-12 spectrometer with a spectral resolution ≈0.2 nm. The obtained data are related to shungite of 98 wt.% carbon.

**Acknowledgement**

A financial support provided by the Ministry of Science and High Education of the Russian Federation grant 2.8223.2013, Basic Research Program, RAS, Earth Sciences Section-5, and grant RFBI 13-03-00422 is highly acknowledged. The authors are grateful to A.Goryunov for assisting in the dispersions preparation.

**References**

1. B. Trauzettel, D.V. Bulaev, D. Loss and G. Burkard. Nat. Phys. **3**, 192(2007).
2. A. Güçlü, P. Potasz and P. Hawrylak. Phys. Rev. B **84**, 035425 (2011).
3. K.A. Ritter and J.W. Lyding. Nat. Mater. **8**, 235 (2009).
4. D. Pan et al. Adv. Mater. **22**, 734 (2010).
5. J. Shen et al. Chem. Commun. **47**, 2580 (2011).
6. Z.Z. Zhang and K. Chang. Phys. Rev. B **77**, 235411 (2008).
7. V. Gupta et al. J. Am. Chem. Soc. **133**, 9960 (2011).
8. R. Liu et al. J. Am. Chem. Soc. **133**, 15221 (2011).
9. Y. Li et al. Adv. Mater. **23**, 776 (2011).
10. T. Zheng et al. ACS Nano **7**, 6278 (2013).
11. L. Tang et al. ACS Nano **6**, (2012).
12. L. Li et al. Nanoscale **5**, 4015 (2013).
13. X. Zhou et al. ACS Nano **6**, 6592 (2012).
14. Y. Dong et al. J. Mater. Chem. **22**, 8764 (2012).
15. M. Zhang et al. J. Mater. Chem. **22**, 7461 (2012).
16. L. Lin and S. Zhang. Chem. Commun. **48**, 10177(2012).
17. S. Chen et al. Chem. Commun. **48**, 763(2012)7.
18. B.S.Razbirin et al. JETP Lett. **87**, 133(2008).
19. E. F. Sheka et al. J. Exp. Theor. Phys. **108**, 738 (2009).
20. B.S.Razbirin et al. Phys. Sol. State 51, 1315 (2009).
21. E. F. Sheka et al. J. Nanophoton. **3**, 033501 (2009).
22. N.N. Rozhkova and E.F. Sheka. arXiv:1308.0794 [cond-mat.mtrl-sci].
23. B.S.Razbirin et al. Advanced Carbon Nanostructures, St.Petersburg, July 1-5, 2013, P. 69.
24. T.A.Witten in Soft Matter Physics. M. Daoud and C.E. Williams (Eds.): (Springer-Verlag, Berlin Heidelberg 1999) P.261.
25. J.-F. Gouyet, Physics and Fractal Structures. Paris/New York: Masson Springer. (1996).
26. S. Park et al. Nano Lett. **9**, 1593 (2009).
27. C. E. Hamilton et al. Nano Lett. **9**, 3460 (2009).
28. N.N. Rozhkova et al. Glass Phys. Chem. **37**, 621 (2011).
29. E.V.Shpol'skii. Physics-Uspekhi **6**, 411 (1963).
30. N.N.Rozhkova, Shungite Nanocarbon. 2011, Petrozavodsk, Karelian Research Centre of RAS (in Russ.).
31. J.-F. Gouyet, (1996). Physics and Fractal Structures. Paris/New York: Masson Springer. 1996.
32. J.P.Heritage, A.M.Glass in Surface Enhanced Raman Scattering. Eds. R.K.Chang and T.E.Furtak. Plenum Press: NY and London. 1982, P.391.
33. E.F.Sheka. Nanosci. Nanothechn. Lett. **3**, 28 (2011).
34. N.N.Rozhkova et al. Smart Nanocomposites **1**, 71 (2010).